\newif\ifUsenix\Usenixtrue %
\newif\ifAnon\Anonfalse

\ifUsenix
  \documentclass[letterpaper,twocolumn,10pt]{article}
  \usepackage{usenix2019_v3}
  \usepackage{authblk}
\else
  \documentclass[conference]{IEEEtran}
\fi
\usepackage{tugraz_defaults}

\usepackage{svg}
\usepackage{listings}
\usepackage{mdframed}
\usepackage[capitalize]{cleveref}
\usepackage{url}
\usepackage{mathtools}
\usepackage{algorithm}
\usepackage{algpseudocode}

\usepackage{caption}
\usepackage{subcaption}

\newcommand{\cc}{\textsc{Cipherfix}}
\newcommand{\cpp}{{C\nolinebreak[4]\hspace{-.05em}\raisebox{.3ex}{\small ++}}}
\newcommand{\ldft}{libdft}
\newcommand{\ldftnew}{libdft64}
\newcommand{\pintool}{Pintool}

\newcommand{\cfbase}{\textsc{Cipherfix-Base}}
\newcommand{\cffast}{\textsc{Cipherfix-Fast}}
\newcommand{\cfenhanced}{\textsc{Cipherfix-Enhanced}}
\newcommand{\xsp}{XorShift128\raisebox{.25ex}{+}}
\newcommand{\rdr}{\texttt{rdrand}}
\newcommand{\aes}{\texttt{AES}}

\newcommand{\encPt}{\mathrm{Enc_{pt}}}
\newcommand{\encMask}{\mathrm{Enc_{mask}}}

\newcommand{\bheading}[1]{{{\textbf{#1}}}}

\renewcommand{\texttt}[1]{$\mathtt{#1}$}

\newcommand{\openssl}{\texttt{OpenSSL}}
\newcommand{\mbedtls}{\texttt{mbedTLS}}
\newcommand{\lsodium}{\texttt{libsodium}}
\newcommand{\wolfssl}{\texttt{WolfSSL}}

\usepackage{color}
\definecolor{halfred}{HTML}{800000}
\definecolor{halfgreen}{HTML}{008000}
\definecolor{codeindent}{HTML}{cccccc}

\newcommand{\repthanks}[1]{\textsuperscript{\ref{#1}}}
\makeatletter
\patchcmd{\maketitle}
  {\def\thanks}
  {\let\repthanks\repthanksunskip\def\thanks}
  {}{}
\patchcmd{\@maketitle}
  {\def\thanks}
  {\let\repthanks\@gobble\def\thanks}
  {}{}
\newcommand\repthanksunskip[1]{\unskip{}}
\makeatother

\begin{document}

\title{Cipherfix: Mitigating Ciphertext Side-Channel Attacks in Software}

\ifAnon  
  \author{Anonymous Submission}
\else
  \ifUsenix    
    \author[]{Jan Wichelmann\thanks{These authors contributed equally to this work.\protect\label{X}}}
    \author[]{Anna Pätschke\repthanks{X}\!\!}
    
    \author[]{Luca Wilke}
    \author[]{Thomas Eisenbarth}
    \affil[]{University of L\"ubeck, L\"ubeck, Germany}
    \affil[]{\textit{\{j.wichelmann,\,a.paetschke,\,l.wilke,\,thomas.eisenbarth\}@uni-luebeck.de}}

  \else
    \author{
      \IEEEauthorblockN{
        Jan Wichelmann\IEEEauthorrefmark{1},
        Anna Pätschke\IEEEauthorrefmark{1},
        Luca Wilke \IEEEauthorrefmark{1},
        Thomas Eisenbarth\IEEEauthorrefmark{1}
    }
    \IEEEauthorblockA{\IEEEauthorrefmark{1}University of L\"ubeck, L\"ubeck, Germany\\
    \{j.wichelmann,a.paetschke,l.wilke,thomas.eisenbarth\}@uni-luebeck.de}
  }
  \fi
\fi

\maketitle

\begin{abstract}

Trusted execution environments (TEEs) provide an environment for running workloads in the cloud without having to trust cloud service providers, by offering additional hardware-assisted security guarantees. However, main memory encryption as a key mechanism to protect against system-level attackers trying to read the TEE's content and physical, off-chip attackers, is insufficient. The recent Cipherleaks attacks infer secret data from TEE-protected implementations by analyzing ciphertext patterns exhibited due to deterministic memory encryption. The underlying vulnerability, dubbed the ciphertext side-channel, is neither protected by state-of-the-art countermeasures like constant-time code nor by hardware fixes.

Thus, in this paper, we present a software-based, drop-in solution that can harden existing binaries such that they can be safely executed under TEEs vulnerable to ciphertext side-channels, without requiring recompilation. We combine taint tracking with both static and dynamic binary instrumentation to find sensitive memory locations, and mitigate the leakage by masking secret data before it gets written to memory. This way, although the memory encryption remains deterministic, we destroy any secret-dependent patterns in encrypted memory. We show that our proof-of-concept implementation protects various constant-time implementations against ciphertext side-channels with reasonable overhead.
\end{abstract}

\section{Introduction}
The current trend for data processing and provisioning of infrastructure heads towards cloud computing, with many co-located clients sharing the same physical hardware instead of working in isolated self-hosted environments.
To protect different clients from each other, as well as the hypervisor from the clients, virtual machines (VMs) are used to provide isolation.
However, especially when processing sensitive data, users may also want isolation from the hypervisor for data privacy or regulative reasons.
This kind of isolation can be provided by trusted execution environments (TEEs), which model the hypervisor as an untrusted party.
To achieve this kind of isolation, TEEs use a combination of additional access rights and cryptography to prevent the hypervisor, or more general, any privileged attacker, from reading the content of the TEE or interfering with its execution state.

Nevertheless, sharing the same hardware leads to traces in shared resources like caches which in turn provides an attack surface for timing or microarchitectural side-channels~\cite{DBLP:conf/crypto/Kocher96,DBLP:journals/cn/BrumleyB05,DBLP:conf/ctrsa/OsvikST06,DBLP:conf/sp/LiuYGHL15,bernstein2005cache}.
A widely used countermeasure against these side-channels is constant-time code that is data oblivious, i.e., does not access memory or decide for branch targets based on secrets~\cite{DBLP:conf/uss/AlmeidaBBDE16,DBLP:conf/ccs/WichelmannSP022}.
To support developers, there are various mostly automated constant-time analysis tools that observe different properties of software traces for finding microarchitectural or timing leakage that could lead to exploitable side-channels~\cite{DBLP:conf/ccs/WichelmannSP022,DBLP:conf/uss/WeiserZSMMS18,DBLP:conf/uss/AlmeidaBBDE16,DBLP:conf/uss/WangWLZW17,DBLP:conf/uss/0011BL0ZW19,DBLP:journals/acm/DanielBR22}.
As these tools advance the constant-time properties of code, leakages get smaller and harder to find, though recent research has shown that even very small
leakages are exploitable, especially when the strong attacker model of TEEs is
considered~\cite{DBLP:conf/ccs/AranhaN0TY20,DBLP:conf/uss/MoghimiBHPS20,DBLP:conf/ccs/SieckBW021}.

The recent Cipherleaks paper~\cite{DBLP:conf/uss/LiZWLC21} and its follow-up~\cite{sytematicCipherleaks} introduced a new
attack vector on code running in TEEs, dubbed the ciphertext side-channel.
The core idea is that some TEEs use deterministic memory encryption, resulting in a \mbox{one-to-one} mapping between plaintexts and ciphertexts for a given memory block. As a result, the attacker
can correlate changes in the ciphertext to the processed data.
For example, the secret decision bit of a constant-time swap operation can be leaked by
observing whether the ciphertext of the corresponding memory location changes,
showing that \mbox{state-of-the-art} constant-time code is not secure under this attacker model.
Thus, this attack vector demands for new analysis methods and countermeasures.

In this work, we introduce an analysis technique to mitigate ciphertext side-channel leakages in constant-time code.
A naive approach hardening every memory write access would result in a very high performance overhead.
Thus, our technique uses secret-tracking to pinpoint critical memory accesses, that are then safeguarded by randomizing observable write patterns such that the resulting binary does not leak information through the ciphertext side-channel.
By combining static and dynamic approaches, we design a solution that covers all program components and works without recompilation.

\subsection{Our Contribution}
We present the \cc{} framework, the first general-purpose drop-in mitigation for ciphertext side-channel-based leakages.
This includes the following contributions:
\begin{itemize}
    \item We propose an analysis technique based on dynamic taint analysis to find
    all secret-containing memory locations in constant-time binaries that are potentially vulnerable to the ciphertext side-channel.
    
    \item We employ dynamic binary analysis to locate stack variables and enable context-aware tracking of heap allocations, in order to support robust static instrumentation.
    
    \item We develop a mitigation technique, based on static binary instrumentation, that hardens the software binary across library boundaries without requiring recompilation and that provides three different security levels.
    
    \item We evaluate our proof-of-concept implementation of \cc{} regarding performance and security on various primitives from four widely-used cryptographic libraries and discuss the  effects of different mitigation approaches. 
\end{itemize}
Our source code is available at \url{https://github.com/UzL-ITS/Cipherfix}.

\vspace{5pt}
\noindent\bheading{Outline.}
After providing background in \Cref{sec:background}, we give an overview over the design of \cc{} in \Cref{sec:design}. In \Cref{sec:analysis}, we present our dynamic analysis, which we use to build the static mitigation as described in \Cref{sec:mitigation}. We evaluate the performance and security of our mitigation in \Cref{sec:evaluation}. Finally, in \Cref{sec:discussion}, we discuss design decisions of \cc{} and point out angles for future work.

\section{Background}\label{sec:background}

\subsection{Secure Encrypted Virtualization}
AMD Secure Encrypted Virtualization (SEV) is a trusted execution environment (TEE) that is designed as a drop-in solution to protect whole virtual machines. It encrypts the RAM content of the VM with an encryption key inaccessible to the hypervisor~\cite{SEV_encryption_doku}.
The latest iteration, SEV Secure Nested Paging (SEV-SNP)~\cite{SEV-SNP_whitepaper},
prevents the hypervisor from remapping or modifying VM memory, thwarting attacks like~\cite{DBLP:conf/sp/WilkeWM020,DBLP:conf/uss/LiZLS19,buhren:2017:fault,du:2017:sevUnsecure,DBLP:conf/eurosys/Morbitzer0HW18}. 
For the memory encryption, SEV uses AES-128 in the XOR-Encrypt-XOR (XEX)~\cite{DBLP:conf/asiacrypt/Rogaway04} mode of operation, where a tweak value is XOR-ed  before and after encryption.
SEV derives the tweak values from the
physical address of a 16-byte memory block and a random seed generated at boot time.

\subsection{Ciphertext Side-Channel}
\label{background:ciphertext-side-channel}
The ciphertext side-channel was first introduced in~\cite{DBLP:conf/uss/LiZWLC21} and later generalized to arbitrary memory regions and implementations in~\cite{sytematicCipherleaks}. 
Both papers extract cryptographic keys from state-of-the-art constant-time cryptographic implementations running in SEV-SNP VMs. While the attack vector in~\cite{DBLP:conf/uss/LiZWLC21} has been fixed on a firmware level~\cite{bulletinVMSAFix}, the attacks from~\cite{sytematicCipherleaks} remain unaddressed.
The core idea is exploiting the deterministic encryption at a fixed memory location, to leak information by precisely observing changes in the ciphertext and correlating them with the (known) executed code.

\begin{figure}[t]
    \centering
    \begin{subfigure}[b]{0.45\textwidth}
        \centering
        \includegraphics[width=\textwidth]{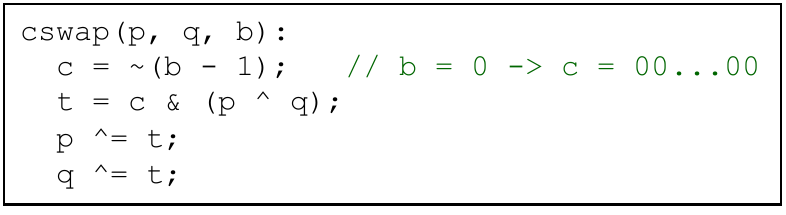}
        \caption{Constant-time swap of \texttt{p} and \texttt{q}, depending on bit \texttt{b}.}
        \label{fig:cswap:code}
    \end{subfigure}
    \par\bigskip
    \begingroup
    \setlength{\tabcolsep}{12pt}
    \begin{subfigure}[b]{0.45\textwidth}
        \centering
        \begin{tabular}{@{\hspace*{1em}}p{8pt}rr@{}}
             & \multicolumn{2}{c}{Ciphertext of \texttt{p}} \\ \cmidrule(lr{-0.5pt}){2-3}
             \texttt{b} & before \texttt{cswap} & after \texttt{cswap} \\
             \midrule
             0 & \texttt{e4c80f2a} & \texttt{e4c80f2a} \\
            1 & \texttt{e4c80f2a} & \texttt{aa2f2a61} \\
        \end{tabular}
        \caption{Ciphertext of \texttt{p}, before and after calling \texttt{cswap}.}
        \label{fig:cswap:table}
    \end{subfigure}
    \endgroup
    \caption{\texttt{cswap} and resulting ciphertexts for the encrypted RAM accessible by the attacker. \ref{fig:cswap:code} shows the procedure of a constant-time swap. Depending on the value of a secret decision bit \texttt{b}, the values \texttt{p} and \texttt{q} are swapped (\texttt{b = 1}), or left as-is (\texttt{b = 0}). \ref{fig:cswap:table} shows the effect on the resulting ciphertext:  If the ciphertext did not change, the attacker can infer that \texttt{b = 0}; if the ciphertext changed, the attacker learns that \texttt{b = 1}.}
    \label{fig:cswap}
\end{figure}

The authors of~\cite{sytematicCipherleaks} introduce two attack variants: The \emph{collision} and the \emph{dictionary} attack. Both attacks exploit repeated write operations to the same memory address. The collision attack extracts information from observing the same ciphertext over multiple writes. One common example is the \texttt{cswap} pattern (\Cref{fig:cswap}): A variable is always written, but depending on a secret decision bit the old or the new value is selected.
While in the former case the deterministic ciphertext remains unchanged, in the latter case a new value is written, producing a different ciphertext.
Thus, by observing the ciphertext of the memory location before and after the \texttt{cswap}, the attacker can immediately infer the secret decision bit.
In the dictionary attack, the attacker does not only rely on collisions, but maps ciphertexts to (partially) known plaintexts. As the dictionary attack relies on repeating ciphertexts as well, mitigating the collision attack also mitigates the dictionary attack.

While the attacks above target values explicitly written to memory by the application, they can also be used to extract register values.
For this, the authors of~\cite{sytematicCipherleaks} exploit that the operating system running in the SEV-protected VM stores the user space register values upon context switches on the stack. %
This mechanism allows an attacker to extract secrets residing in registers by forcing context switches and observing the ciphertexts. However, the authors also describe how to fix this issue, by randomizing the stack layout. %

\subsection{Binary Instrumentation}
\label{sec:background:instrumentation}
Binary instrumentation allows modifying compiled programs without access to the source code. This is commonly used to insert new code that gathers information.

\bheading{Dynamic binary instrumentation} (DBI) gives the opportunity to include the architectural state by executing the analysis routines while the program is running. There are numerous DBI frameworks, e.g., Valgrind ~\cite{DBLP:conf/pldi/NethercoteS07}, Intel Pin~\cite{DBLP:conf/pldi/LukCMPKLWRH05}, DynamoRIO~\cite{DBLP:conf/cgo/BrueningGA03} or DynInst~\cite{DBLP:journals/ijhpca/BuckH00}.
The Intel Pin framework compiles and inserts analysis instructions at runtime through an x86 just-in-time (JIT) compiler.
The code is processed in units called \emph{basic blocks}, which are defined as instruction sequences that have a single entry and exit point.
Through a number of callbacks, a so-called \emph{\pintool{}} specifies the analysis code to be inserted during JIT compilation.
The original instructions and the analysis code are combined such that the instrumentation is transparent to the analyzed program.

\bheading{Static binary instrumentation} (SBI) results in a modified standalone binary that is obtained by the use of rewriting or redirecting techniques.
The execution of an instrumented binary does not depend on an instrumentation framework, which means that the main overhead comes from the inserted analysis code~\cite{andriesse2018practical}.
However, static instrumentation struggles with analyzing indirect branches, shared libraries and dynamically generated code~\cite{DBLP:conf/ispass/LaurenzanoTCS10,DBLP:conf/pldi/LukCMPKLWRH05}.
There are different approaches for adding analysis code to the binary at specific instrumentation points and then redirecting the control flow, such that both analysis and original application code are executed in the right order. To avoid breaking references, the instrumented code can be put into a separate \texttt{.instrument} section. %
An instrumentation point then redirects execution to this section,
either through software breakpoints via the \texttt{int3}~\cite{DBLP:conf/cgo/NandaLLC06} instruction and a custom signal handler, or through direct jumps via so-called \emph{trampolines}~\cite{DBLP:journals/ijhpca/BuckH00,DBLP:conf/IEEEpact/HollingsworthMGNXZ97,hunt1999detours}.
It is possible to combine multiple approaches to minimize their shortcomings,
e.g., by inserting 5-byte jumps
where possible, and falling back to 2-byte jumps
or \texttt{int3} when not enough space is available.
An example of trampoline-based instrumentation is illustrated in \Cref{fig:static-instrumentation} in the appendix.
Recent binary rewriting approaches further optimize the instrumentation through using available metadata for lifting~\cite{DBLP:conf/asplos/Williams-KingKW20} or symbolization of references~\cite{DBLP:conf/sp/DineshBXP20}.

\subsection{Dynamic Taint Analysis}
Dynamic taint analysis (DTA) tracks the flow of selected information through a program during code execution.
The data to be tracked is marked as a \emph{taint source}, and its propagation is defined through a \emph{taint policy}.
The policy also determines 
the \emph{taint sinks} that can be reached by the data.
All instructions that process secret data are considered for the taint propagation.
Data flow tracking can be done in various granularities, whereby byte-level tracking is the most commonly used.
For each memory location and register, there is shadow memory containing the taint label information, so the performance overhead is directly connected to the granularity. If too much data is marked as tainted, this is called \emph{overtainting}; tainting too little data is referred to as \emph{undertainting}~\cite{DBLP:conf/vee/KemerlisPJK12,andriesse2018practical,DBLP:conf/sp/SchwartzAB10}. 

A widely-used x86 taint analysis tool providing fast taint propagation based on Intel Pin is \ldft{}~\cite{DBLP:conf/vee/KemerlisPJK12}.
In order to also support 64-bit binaries, \ldft{} has been extended for VUzzer64~\cite{DBLP:conf/ndss/0001JKCGB17} and the AngoraFuzzer~\cite{DBLP:conf/sp/ChenC18}.
The data flow-based byte-level taint propagation in \ldftnew{} is implemented through handwritten rules for every instruction class. %

\section{\cc{} Design}
\label{sec:design}
We first give an overview of the generic design of our ciphertext side-channel countermeasure.

\subsection{Attacker Model}
\label{sec:design:attackermodel}
We assume an attacker that tries to extract secret information from a TEE, that is protected with a deterministic block-based memory encryption with address-dependent tweaks. The attacker knows the exact binary which is executed by the victim, but cannot access secret data that is stored within the TEE. They have root access to the machine running the TEE and are able to read the entire encrypted memory, but cannot decrypt or modify it. Furthermore, the attacker can make use of a controlled channel that allows them to track and interrupt the code running inside the victim's TEE.
This means that they can reconstruct the entire control flow of the targeted application and annotate it with snapshots of the corresponding ciphertexts in memory. One instance of such a scenario is a malicious hypervisor attacking a VM that is protected with AMD SEV-SNP.
Finally, we assume that potential operating systems running alongside the
targeted application inside the TEE do properly protect register values from ciphertext side-channels attacks, 
as discussed in \Cref{background:ciphertext-side-channel}.%

\subsection{Countermeasure Requirements}\label{sec:design:requirements}
Our overall goal is to produce a hardened binary which does not contain leaking memory writes. The countermeasure should not only protect the targeted program itself, but all its dependencies as well, as leakage may span multiple libraries (e.g., a crypto library calls \texttt{memcpy} in libc), and library developers are unlikely to widely adopt ciphertext side-channel countermeasures themselves. Finally, we target application developers who build code on top of third-party libraries and who do not have the necessary insight to manually fix leakages in those libraries. Thus, a drop-in solution with little manual interaction is desirable here.

There are two major approaches to this: One could either create a compiler extension that rewrites vulnerable memory accesses at compile time, or modify existing binaries through SBI.
A pure compiler-based solution needs to recompile all dependencies, which is complex and requires manual intervention.
A combination of DBI and SBI can work directly with the compiled binaries and, given sufficient coverage, accurately identify and harden vulnerable memory writes.
For these reasons, \cc{} aims for a binary instrumentation-based solution.
The trade-off between binary vs. source-based approaches is further discussed in \Cref{sec:discussion:code-based-instrumentation}.

\subsection{Protecting Memory Writes}\label{sec:design:protectingwrites}

In order to protect an existing binary from being attacked through a ciphertext side-channel, the content-based patterns of write accesses to memory have to be obscured. In~\cite{sytematicCipherleaks}, the authors propose various approaches for randomizing observed ciphertexts: First, by limiting reuse of memory locations through using a new address for each memory write; second, by interleaving data with random nonces; and third, by applying a random mask when writing data.
The first approach uses the fact that different memory addresses get different tweak values in the memory encryption, but has a high overhead when applied outside of well-defined conditions. The second approach requires extensive changes to data structures, which has many pitfalls and needs to be done by the compiler. Due to lower overhead and higher practicability, we thus opt for the last approach, i.e., we add a random mask whenever an instruction writes secret data to main memory.
We further discuss the different approaches in \Cref{sec:discussion:masking-alternatives}.

The masking of data takes place before memory writes and after memory reads. To store the masks belonging to a particular memory chunk (e.g., a \cpp{} object), we allocate a \emph{mask buffer} of the same size, so there is a one-to-one mapping of data bytes to mask bytes. When writing data, we generate and store a new mask, XOR it with the plaintext, and store the masked plaintext; when reading, we read the mask and then decode the masked plaintext. Note that we need to ensure that at no point non-encoded secret data is written to memory, so all decoding must be done in secure locations like registers.

\subsection{Tracking Data Secrecy at Runtime}\label{sec:design:tracking-secrecy}%
While masking all memory writes provides good protection, it comes with a high overhead. In fact, only a fraction of all memory writes relate to secret information: As we assume that the implementation is constant-time, there is no secret-dependent control flow, so, for example, return addresses pushed onto the stack by function calls can be safely written in clear text. The same is true for the data structures used by the heap memory allocator to keep track of memory chunks. Finally, there may be a point where data is no longer considered secret, e.g., when sending a signature over the network. We thus aim to find and protect those instructions that actually deal with secret data. However, this is non-trivial, as there may be instructions that access both public and secret data, depending on the context (e.g., from \texttt{memcpy}).

\begin{figure}
    \centering
    \begin{subfigure}[t]{0.23\textwidth}
        \centering
        \includegraphics[width=\textwidth]{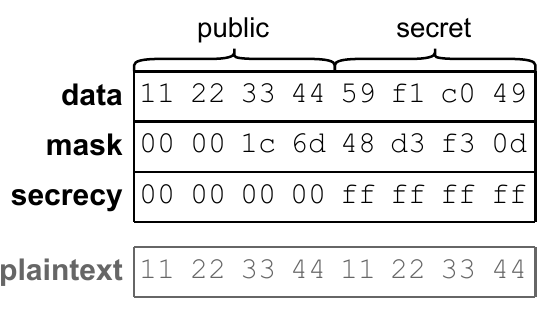}
        \caption{\cfbase{}}
        \label{fig:secrecy:base}
    \end{subfigure}
    \begin{subfigure}[t]{0.23\textwidth}
        \centering
        \includegraphics[width=\textwidth]{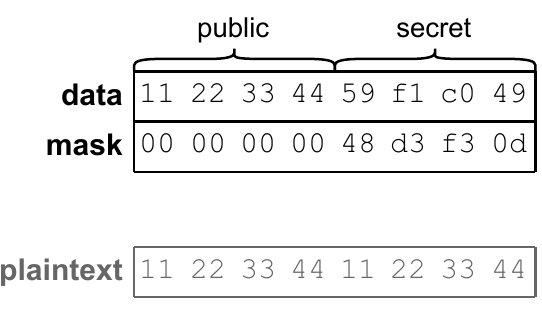}
        \caption{\cffast{}}
        \label{fig:secrecy:fast}
    \end{subfigure}
    \caption{\cfbase{} stores the secrecy information in a separate buffer, and uses it to decide whether a given mask byte should be applied or not. This allows to safely have non-zero mask bytes behind public data, as they are ignored if the corresponding secrecy bytes are zero. In contrast, \cffast{} stores this information directly in the mask buffer, i.e., a mask byte is zero iff the corresponding data is public.}
    \label{fig:secrecy}
\end{figure}

Thus, we need a way to detect at runtime whether a given memory address should be considered secret, i.e., whether the data at that address is masked, and whether we should apply a new mask when writing to said address. We propose two approaches for storing this \emph{secrecy} information (Figure~\ref{fig:secrecy}): In the first approach, which we denote \cfbase{}, we allocate another buffer of the same size as the mask buffer, called the \emph{secrecy buffer}. In the second approach, \cffast{}, we encode this information directly into the mask buffer.

\subsubsection{Storing secrecy information separately}\label{sec:design:cfbase}
In \cfbase{} we allocate a buffer that holds the secrecy information for each memory location. If a byte is public, the corresponding secrecy byte is \texttt{0x00}; if a byte is secret, the secrecy byte is \texttt{0xff}. The secrecy buffer is initialized on allocation, and may be updated during the lifetime of the object. This construction allows us to read and update data without branching, as we can combine the secrecy value $S$ with the mask $M$ via a bitwise AND ($\otimes$), before applying it to the data via a bitwise XOR ($\oplus$): When reading, we compute $P=\hat{P}\oplus (M\otimes S)$, so we only decode the stored (potentially masked) plaintext $\hat{P}$ if the address is considered secret. For writing, we always generate and store a new mask, and then compute $\hat{P}=P\oplus (M\otimes S)$ for plaintext $P$.
As we make no assumptions about the mask, this generally functions as a one-time pad: The mask $M$ is fully random and independent from the plaintext $P$, thus $\hat{P}$ is independent from $P$ as well. 

\subsubsection{Storing secrecy as zero masks}\label{sec:design:cffast}
By separating mask and secrecy information, \cfbase{} can generate uniform masks, yielding a one-time pad encoding. However, this comes at a cost: First, we get high memory overhead by allocating the mask and secrecy buffers. Second, each read is replaced by three reads, namely to the data, mask and secrecy buffers. To reduce this overhead, we make an observation: If the data is public, ANDing the mask and the secrecy value yields zero; if the data is secret, we use the mask value directly. Thus, for \cffast{}, we merge the secrecy information and the mask into the mask buffer, by setting the mask to zero when the data is public, and to a random non-zero value otherwise.

For writes, we check whether the old mask is zero before generating a new one, saving a memory write in some cases; for reads, we directly XOR the mask value, saving a memory read compared to \cfbase{}. Thus, in addition to the reduced memory overhead, we get a performance improvement due to fewer memory accesses.
We discuss the security implications of this in \Cref{sec:eval:security}.

\subsubsection{Reducing risk of mask collision}\label{sec:design:cfenhanced}
\begin{figure}
    \centering
    \includegraphics[width=0.48\textwidth]{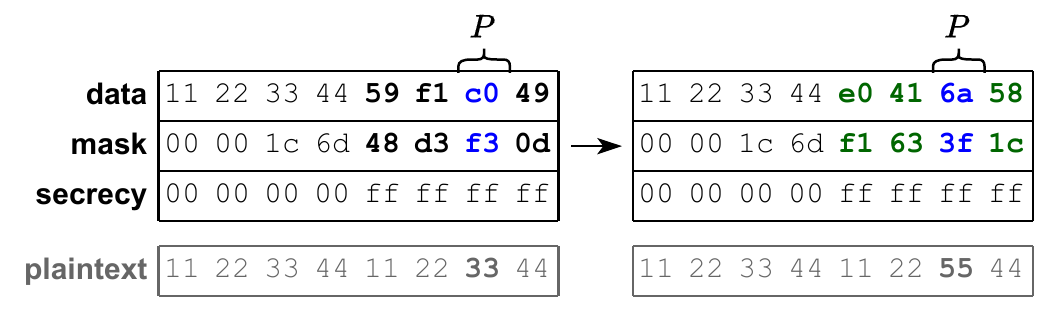}
    \caption{Extended write in \cfenhanced{}. Instead of updating only $w=8$ data and mask bits at offset 6, \cfenhanced{} extends the write to $w'=32$ bits, by also updating the mask of the surrounding three bytes, reducing the probability of a mask collision.}
    \label{fig:cfenhanced}
\end{figure}

While \cc{} can be used with secret data of any size, the width of the masks influences the robustness against attackers that observe ciphertexts over longer periods of time. For example, for a $w=8$ bit wide mask, a mask collision can be expected in as few as $\sqrt{2^w}=16$ writes. To address this issue, we propose \cfenhanced{}, which, as an extension of \cfbase{}, converts writes with a size $w$ below a certain threshold to a bigger size $w'$ that is considered safe: Instead of updating $w$ bits, we generate a new mask of size $w'$ bits and update $w'$ data bits at once. This is possible due to architectures like x86 supporting multiple write sizes from 1 byte to 8 bytes (and even more with vector instructions). We thus read and decode the existing masked plaintext $\hat{P'}$ around the given address, merge it with the new plaintext $P$ and then re-encode it. A write access protected with \cfenhanced{} is illustrated in \Cref{fig:cfenhanced}.

\subsection{Toolchain}\label{sec:design:toolchain}
\begin{figure}
  \centering
  \includegraphics[width=\linewidth]{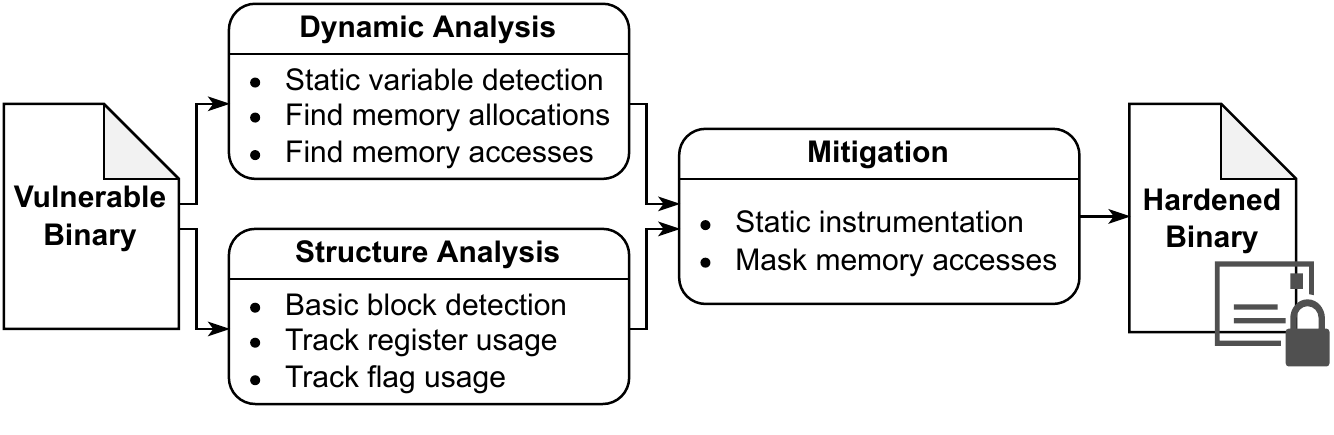}
  \caption{Structure of the \cc{} framework. The vulnerable binary is dynamically analyzed and then hardened through static instrumentation.}
  \label{fig:framework}
\end{figure}

The \cc{} framework is a drop-in solution that analyzes existing binaries with DTA to identify vulnerable code and then statically instruments the binaries to mitigate the detected leakages.
\cc{} consists of two distinct steps (\Cref{fig:framework}). In the analysis step, a taint analysis tool detects instructions and memory locations like stack frames and heap objects, that touch secret data. In parallel, a structure analysis tool extracts information about basic blocks and register/flag usage per instruction to aid the static mitigation. Finally, the mitigation step uses the analysis results to statically instrument the vulnerable binaries, inserting masking code for secret memory accesses and installing infrastructure for initializing newly allocated memory. In the following sections, we discuss the respective steps in more detail.

\section{Leakage Localization and Preprocessing}
\label{sec:analysis}

In order to protect read/write operations, we first need to identify all vulnerable memory locations and the instructions accessing them. Our static mitigation relies on some additional structural information, i.e., the offsets of basic blocks and liveness of registers and flags.
In the following, we describe our leakage localization technique and the other analysis steps.

\subsection{Dynamic Secret Tracking}
With the help of DBI and DTA, we can collect information that is only available at runtime.
As constant-time code does not include secret-dependent control-flow, DTA covers all paths of the implementation.
For the cases of non-constant control flow in public paths, we use multiple iterations of the program with different inputs.
We further discuss this in \Cref{sec:discussion:non-constant-time}.
If an exact analysis is not possible, we stay on the safe side and avoid undertaining so that in combination with full path coverage we reliably identify all secret accesses.

Our proof-of-concept implementation is based on \ldftnew{} data flow tracking.
When combined with a Boolean taint, we found that byte-level tainting of memory is fast enough to analyze complex cryptographic libraries while maintaining a high accuracy. While that leads to some overtainting (i.e., some memory locations get protected unnecessarily), we avoid undertainting. 
We also extended \ldftnew{} by adding support for many SSE/AVX vector instructions, which are heavily used in optimized cryptographic code. 
All in all, we added \num{4355} lines of code (LoC) to \ldftnew{} for new instruction support and \num{2269} LoC for our tracking logic.%

\subsubsection{Taint policy}
We offer several venues for specifying taint sources, depending on the use case: First, if the main application itself can be easily recompiled (e.g., a custom network program linked against OpenSSL), the developer can call a special \texttt{classify} function, which takes a memory address and a size parameter. The taint analysis \pintool{} tracks this function and introduces taint for the corresponding memory when observing a call. In addition, we support fully automated assigning of taint sources without recompilation: Many cryptographic implementations read their private keys from the file system, so by intercepting the \texttt{open} and \texttt{read} system calls we can detect accesses to such files and taint the incoming data.

Our policy does not introduce taint sinks in the classical way; instead, those consist of all traced memory accesses and information that is needed for the static countermeasure.
However, we offer a \texttt{declassify} function that explicitly marks data as no longer secret, i.e., all associated taint is deleted. In addition, functions that transmit data over insecure channels (e.g., network functions) remove taint as well.
Thereby, we ensure that data that is meant to be publicly available does not get damaged by remaining secrecy features.

\subsubsection{Tracking secret-related instructions}
In order to protect memory accesses in our mitigation, we need to identify all instructions that read or write secret data at some point of the execution.
The analysis distinguishes between three different cases: For instructions that only process public data there is no need to apply any ciphertext side-channel protection, whereas for instructions that only process private data the content written to memory always gets randomized. Finally, there are instructions that only occasionally access secrets and thus need to be able to distinguish between public and secret memory. As the latter may come with a certain performance overhead, the information about secrecy of accessed memory should be included in the taint analysis result used for the static mitigation.

\subsection{Identifying Memory Locations}
As the taint analysis itself tracks secrets only through ``raw'' memory addresses, it misses a lot of context: For example, there is no distinguishing between heap and stack memory, and which function a given accessed stack frame belongs to. However, for a static mitigation, we need certain information about each object in memory, like where it is allocated and which offsets need to be protected.
There are various kinds of memory locations, i.e., static variables in the binary itself and dynamically allocated heap blocks and stack frames, so we need to distinguish between those cases.

\subsubsection{Finding static variables}
\label{sec:analysis:image-block-detection}
During the execution of cryptographic code, some instructions access data that lies within the memory region of the mapped binary, i.e., static initialized or uninitialized variables. Since we cannot access source-code level information about the program, we develop a method to locate these variables and determine their size, as we aim to only protect those that  contain secret information. 
These fine-grained memory objects keep their secrecy status during the whole execution (i.e., if a variable contains secrets at some point, it is secret from the beginning until the end of a program run).
For the static variable detection, we implemented a small \pintool{} with \num{338} LoC that collects traces of memory accesses in the data segments, matches these accesses to contiguous blocks in the binary's memory region and then produces an output file that can be parsed by the main taint tracking \pintool{}.

\subsubsection{Heap allocations}
\label{sec:analysis:heap}
\begin{figure}[t]
    \centering
    \includegraphics[width=0.43\textwidth]{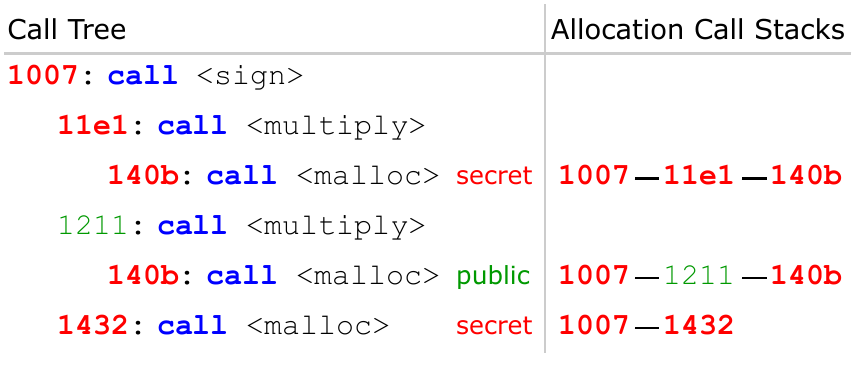}
    \caption{Call tree and resulting call stacks for three heap allocations. The call stack is accompanied with secrecy information, i.e., whether a secret block was allocated. The offsets of call instructions that lead to at least one secret allocation are marked bold and red; the offsets of instructions that only lead to public allocations are marked green. This information is directly reused in the instrumentation (see \Cref{sec:mitigation:allocations:heap}).}
    \label{fig:allocation-analysis}
\end{figure}

Heap allocations are tracked through explicit (de)allocations, e.g., the \texttt{malloc}, \texttt{realloc}, \texttt{calloc} and \texttt{free} standard library functions.
Similar to the static variables, the secrecy status of a heap object is kept for its entire lifetime. However, the heap layout may be different for each execution, so we cannot rely on fixed addresses to identify a heap object. Generating a flat list of heap allocations and retrieving the secrecy information using a counter variable is not useful either, as this restricts the hardened binary to a single control flow path. Instead, we use the call stacks of the heap allocations: Apart from rare cases where allocations are done in a loop, the call stack of each allocation is unique and thus suitable for identifying it both during analysis and at runtime. The call stack of an allocation is determined by keeping track of all calls and returns during analysis and emitting the current call stack whenever an allocation function is observed (\Cref{fig:allocation-analysis}).

As for heap objects the application itself has full control over their layout and the stored data types may vary depending on context, we cannot safely make assumptions about relative offsets within a heap object. Thus, we opted for marking the entire object as secret whenever a part of it gets tainted. While this overapproximation may lead to a slightly higher overhead due to protecting more instructions than strictly necessary, it reduces complexity and makes the static mitigation more robust. We also found that in practice the impact is limited, as generally the size of a heap object correlates with the amount of (private) data stored in it (e.g., big integer objects).

\subsubsection{Tracking stack frames}
\label{sec:analysis:stack}
The stack memory area is characterized by rather liberal (de)allocation and access strategies, which makes separating individual stack frames difficult and thus complicates tracking the exact offsets and lifetimes of secret variables. An easy solution would be marking the entire stack as secret and protecting all instructions that ever access stack memory, but this would introduce a lot of unnecessary overhead, since the stack is mostly used for temporarily storing registers and small local variables that often do not contain secret data.
Instead, in order to avoid overtainting and the aforementioned performance penalty, we developed a generic stack frame tracking strategy that allows to keep track of secret data throughout the program execution by means of stack frame offsets.
Contrary to the heap, the stack usually conforms to fixed patterns built by the compiler, so we can assume that relative offsets within a function's stack frame are valid over multiple executions.

Our proof-of-concept implementation does not rely on source code or function symbols, but works with any standard-conforming binary.
The stack allocation tracking consists of identifying function calls, mapping a call target to an actual function
for which a stack frame initialization of the static instrumentation is needed
and determining its respective stack frame size, and building a list of secret offsets within that stack frame.

Most function calls are detected through \texttt{call}/\texttt{ret}-pairs; in addition, our analysis includes a heuristic for detecting tail calls, i.e., when a function is exited via a \texttt{jmp} instruction to another function. Calls to functions in shared libraries present another challenge, as the application invokes those through a call to the \texttt{.plt} section, which may in turn jump into the dynamic runtime linker to resolve the actual function call target.
In order to find the function in the shared library and not its stub code in the caller's \texttt{.plt} section, we need to follow the resolution process in the dynamic linker until we reach the actual call target.
This is done through a state machine that keeps track of the current linking state and generates a mapping of \texttt{.plt} offsets to the corresponding functions.

After detecting a function, we proceed with determining its stack frame size. This is achieved through several means: First, there may be explicit stack frame allocations through instructions like \texttt{push}/\texttt{pop} and \texttt{sub}/\texttt{add}, which directly modify the stack pointer. In addition, the x86-64 ABI permits functions to freely use a small chunk above the stack pointer (which usually marks the end of a stack frame), the so-called red zone. We handle this by updating the stack frame size whenever we observe an access outside a known stack frame.

\subsection{Binary Structure Analysis}
\label{sec:analysis:structure-analysis}
Contrary to DBI, where the executed code is recorded and instrumented at runtime, SBI must apply all changes in an offline manner, without being able to handle unexpected states. 
Our proof-of-concept SBI-based mitigation needs further information besides the DTA results, namely the precise bounds of all basic blocks and, for each instruction, the usage of registers and status flags. The latter is necessary since the masking operations need scratch registers to store intermediate results, and inadvertently clobber the status flags. While this information can be collected through static liveness analysis or heuristics~\cite{DBLP:conf/sp/DineshBXP20,DBLP:conf/asplos/Williams-KingKW20}, we decided to employ dynamic analysis here as well, as we already have the necessary code coverage from the DTA. This approach marks only registers and flags that are indeed used, avoiding unnecessary saves/restores and thus reducing the runtime overhead. We created a specialized \pintool{} with \num{599} LoC that collects the aforementioned information and passes it to the SBI tool.

\section{Static Mitigation}
\label{sec:mitigation}
With the information from the dynamic analysis we can now statically instrument the affected binaries, hardening them against ciphertext side-channel attacks.
We identify consecutive basic block chains (functions), which are then copied and instrumented at a new section in the binary. The original code locations are replaced by a number of jumps to their instrumented counterparts, following an optimized trampoline-approach described in \Cref{sec:background:instrumentation}.
We then modify all vulnerable memory accesses to apply masks, such that each of these memory writes is randomized.
The resulting hardened binaries are self-contained, i.e., they can be executed without an external instrumentation framework.

\subsection{Masking Memory Accesses}
\label{sec:mitigation:masking}
\begin{figure}[t]
    \centering
    \includegraphics[width=0.48\textwidth]{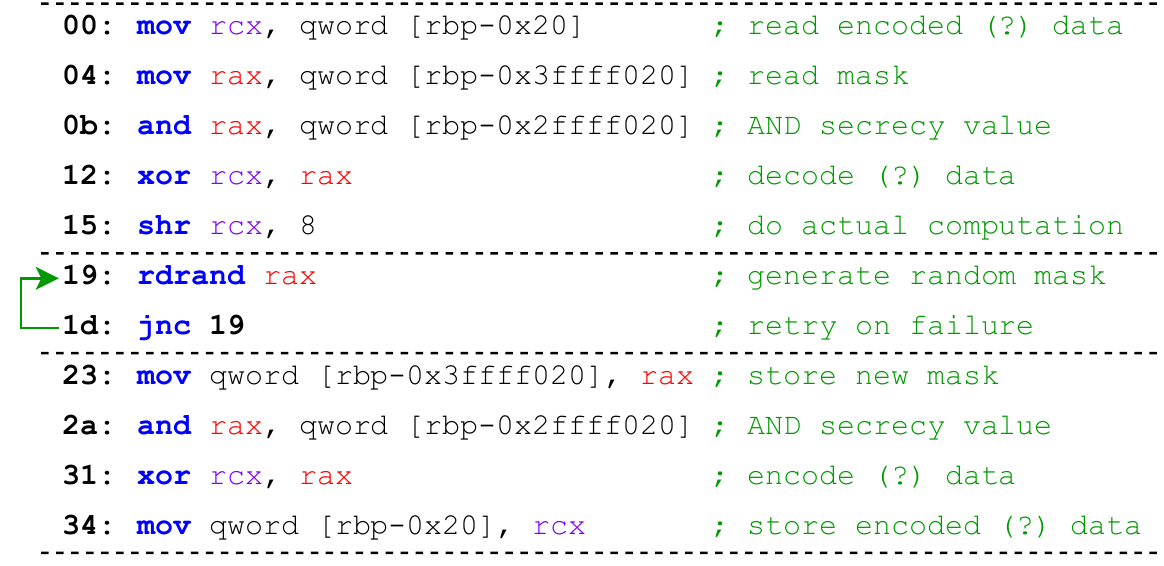}
    \caption{Assembly code generated by \cfbase{} for the instruction \texttt{shr~qword~[rbp-0x20],~8}, that accesses both public and secret memory. As an in-place shift, it has to first read and decode the left operand, compute the shift, and then encode and store the result. The instrumentation tool identified \texttt{rax} and \texttt{rcx} as scratch registers, which did not need to be preserved.}
    \label{fig:instrumented-shr}
\end{figure}

After copying all affected basic blocks to a separate section, we can replace the vulnerable memory accesses by hardened instruction sequences.
As described in \Cref{sec:design:tracking-secrecy}, we mitigate the ciphertext side-channel by adding a random mask to each memory write to a secret location.
Some instructions have read and write accesses (e.g., arithmetic with a memory operand acting both as source and destination), so they may need decoding and encoding (\Cref{fig:instrumented-shr}). String operations like \texttt{rep~movsq} are replaced by an explicit loop that decodes each word of the source data and re-encodes it for the destination, as not the entire copied memory block may be secret. Our proof-of-concept implementation supports protection of common arithmetic and move instructions, and a number of vector instructions that occur in cryptographic code. %

Each memory block is accompanied by a mask buffer and a secrecy buffer, which have a constant distance $d_M$ resp. $d_S$ to the memory block's address. Using a constant distance saves expensive look-ups for finding the appropriate buffers, reducing the total overhead of the mitigation. For our test setup, we found that $d_M=\mathtt{0x3ffff000}$ and $d_S=\mathtt{0x2ffff000}$ work well. These provide sufficient memory space while still fitting into the signed 32-bit memory displacement immediate which is supported by x86-64, and avoid penalties like aliasing when two addresses share too many low bits.

\subsubsection{Updating the masks}
\label{sec:mitigation:updating-the-masks}
Apart from initializing the mask and/or secrecy buffers during setup (see \Cref{sec:mitigation:allocations}), we need to update the mask values before every write operation.
To ensure that masks do not have repeating or easily exploitable patterns, we sample them from a pseudorandom number generator (PRNG). As we want to keep the overhead low, any such PRNG should have a small code footprint and require as few registers as possible, which rules out most classic software-based PRNGs. A natural choice on x86-64 is the \rdr{} instruction, which fills a single general purpose register with random bytes.
The instruction offers cryptographically secure randomness. %
However, its security guarantees also lead to a noticeable slowdown when the instruction is used extensively.

To work around this, we devised two additional PRNGs for mask generation. The first one, named \aes{}, makes use of the AES-NI \texttt{vaesenc} instruction to repeatedly apply the first round of AES to an initially random 16-byte state with a random 16-byte round key.
The second PRNG is \xsp{}, a widely-used and fast full-period generator~\cite{DBLP:journals/jcam/Vigna17}, for which we created a vectorized implementation.
In both cases, the new mask is extracted from the state.
For best performance, the \aes{} PRNG needs two vector registers and the \xsp{} PRNG needs three. We found that usually enough such registers are available, and, if not, the overhead for the additional save/restore is still smaller than calling \rdr{}.
We discuss the properties of the different PRNGs in \Cref{sec:eval:leakage-sources}.

\subsubsection{Scratch registers and flags}
For some operations, we need additional scratch register space for storing intermediate results. Since we are restricted to working with an existing binary, we cannot exclude registers from being allocated by the compiler and thus have to look for registers which hold stale values, or save those values in a secure location. We use the results from the structure analysis in \Cref{sec:analysis:structure-analysis} to identify suitable registers. To save general purpose registers, we prefer using SSE vector registers via the \texttt{vmovq} and \texttt{vpinsrq} instructions, as those are fast and immune to ciphertext side-channels. In the rare case where no vector register is available, we store the scratch register's original value in memory. To avoid the expensive masking when writing a secret value to memory, we prioritize registers that the taint tracking did identify as not holding secret data.

Similar to the registers, our instrumentation may overwrite status flags through the encoding/decoding instructions.
To save and restore single flags, we use the \texttt{set\textit{cc}} instruction family, while for multiple flags we rely on the \texttt{lahf} instruction, which copies the entire flag state into the \texttt{ah} register.

\subsection{Managing Mask and Secrecy Buffers}
\label{sec:mitigation:allocations}
The instrumented instructions assume that there is a mask buffer and a secrecy buffer with a constant distance to the accessed memory address. Thus, for each memory block that is accessed by such an instruction, we need to allocate a mask buffer at the corresponding address and initialize it with random data, if it contains secret data.
This comes with a few challenges: First, there are several ways of allocating memory, namely the stack, the heap and static fixed-size arrays in the binary itself. Then, not all memory blocks in these regions are considered secret, so their masks and secrecy values need to be initialized context-aware.
In the following, we discuss strategies for handling the various memory regions.

\subsubsection{Stack}
\label{sec:mitigation:allocations:stack}
The stack is allocated by the operating system at application start and is used for storing return addresses, register values and small local variables. The taint analysis produces stack frame information for each function, which contains the size of the stack frame and the relative offsets where secret data is stored. Accordingly, we insert a small code gadget at the beginning of each function, that prepares its stack frame by generating a random mask or setting the secrecy value for the respective offsets.
The mask and secrecy buffers for the stack are allocated on startup; the constant buffer distances work well for the stack, as it usually resides within a well-known memory range and does not grow beyond a few megabytes.

\subsubsection{Heap}
\label{sec:mitigation:allocations:heap}
For most Linux applications, the heap is a contiguous memory region that is managed by the standard library's allocator. The heap starts at a random base address, and is resized via the \texttt{brk} system call. The user then typically allocates memory by calling \texttt{malloc} or \texttt{realloc}, which ensure that enough heap memory is available and return an appropriate memory range.

\begin{figure}[t]
    \centering
    \includegraphics[width=0.43\textwidth]{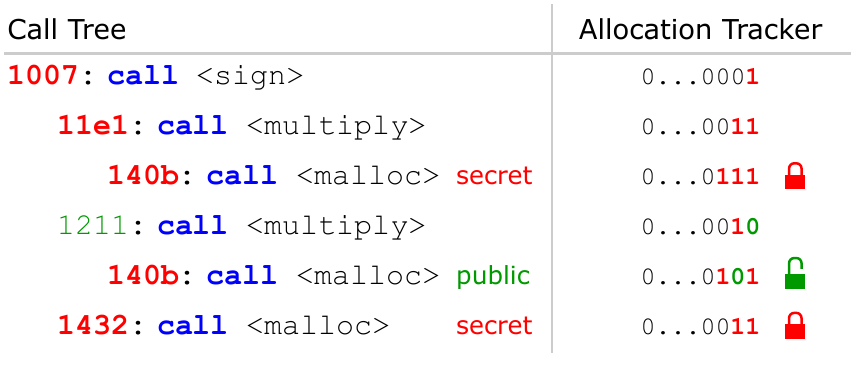}
    \caption{Allocation tracking for the example from \Cref{fig:allocation-analysis}. Each time a \texttt{call} instruction is executed, the allocation tracker is shifted to the left, and 1 is added when this particular \texttt{call} is part of a call tree leading to an allocation of a secret heap object. On return, the tracker is shifted back to the right. The \texttt{malloc}/\texttt{realloc} handler code then checks whether the allocation tracker has the value $2^n-1$, i.e., whether it is all ones starting with the least significant bit. In this case, the new heap object is considered secret; else, it is public.}
    \label{fig:allocation-tracking}
\end{figure}

To guarantee that there are mask and secrecy buffers backing the entire heap region, we instrument the \texttt{brk} system call and (de)allocate corresponding memory each time the heap grows or shrinks. The buffers are initially set to zero.
We also replace the \texttt{malloc} and \texttt{realloc} calls by custom code, which ensures that the corresponding mask and secrecy buffers are correctly initialized depending on whether the allocated memory should contain secret data or not. To identify the particular heap allocation, we resort to tracking its call stack, as explained in \Cref{sec:analysis:heap}. 
We achieve this through an \emph{allocation tracker}, which is an integer residing at a fixed memory address, and which is updated on each \texttt{call} instruction that is part of a call stack that leads to a heap allocation.
Before each \texttt{call}, we left-shift the tracker variable, and add 1 if the call is part of a call stack that leads to allocation of a secret heap memory object.
With our allocation tracker, we can reliably handle heap allocations even if we encounter non-constant control flow or when a function is reused in a different context.
An example is illustrated in \Cref{fig:allocation-tracking}.

Contrary to \texttt{malloc}, the \texttt{realloc} function allows resizing or reallocating an existing heap memory object, while keeping its contents. As the new object may have a different secrecy setting than the old one, we have to ensure that the data is correctly decoded, copied and encoded. However, \texttt{realloc} itself is not aware of the masks and secrecy information, so to avoid losing information, our \texttt{realloc} handler copies the old data, mask and secrecy buffers to a separate memory location, runs \texttt{realloc}, and then restores the contents at the new location with the appropriate encoding.%

If the instrumented program allocates lots of memory, the constant distance to the mask and secrecy buffers may be insufficient, as the heap could at some point overlap with its mask buffers. In this case, one could replace the affected \texttt{malloc} calls by a custom allocator, that is injected into the instrumentation and operates outside the usual heap area. %
Note that this still limits the maximum memory object size to the distance between a memory address and its buffers, i.e., at most two gigabytes, if the instrumentation should do without another scratch register for computing larger offsets.

\subsubsection{Static arrays}
\label{sec:mitigation:allocations:image}
Finally, the binary may have a number of static global variables, which reside in its data sections.
We embed the information about static memory objects containing secret data in the instrumented binary. On startup, an initialization routine walks through this list and allocates and initializes the respective mask and secrecy buffers.

\subsection{Implementation}
We created a proof-of-concept implementation of our mitigation in C\#, which takes the dynamic analysis results and the target program and produces statically instrumented binaries.
The instrumentation tool has \num{7346} LoC, which includes a specifically developed library for patching ELF64 files. 

\subsubsection{Instruction instrumentation}
The instrumentation tool loads and parses the outputs from the taint tracking and the structure analysis tools, and decodes the target ELF files. Then, for each individual binary, the instrumentation
is applied: First, we
look for contiguous basic block chains and 
identify appropriate code locations for inserting jumps to instrumentation code.
Next, we replace each memory accessing instruction marked by the DTA by a masked version. %
After handling all basic blocks, we obtain a list of unmodified and instrumented instructions, grouped by their respective basic blocks.
In a final step, we re-assemble those instructions and write them into a newly allocated ELF section, while patching the basic blocks in the old \texttt{.text} section to jump to the instrumentation code.

\subsubsection{Initialization}
After the instruction-level instrumentation is done, we need to install infrastructure for handling the \texttt{int3} signals and some initialization code that allocates mask and secrecy buffers. For this, we created an \emph{instrumentation header}, which consists of \num{966} lines of assembly code interleaved with some static constants which are later replaced by the instrumentation tool. The instrumentation header hooks into the \emph{constructor} of each binary, which is executed by the dynamic linker when a binary is loaded into memory. This way, we ensure that our initialization runs before all other application code. The initializer of the main program sets up the signal handler, and determines the stack size and base address. Then, it allocates mask and secrecy buffers for the stack. The initializers of the main program and of all dynamic libraries iterate through the list of secret static variables deposited by the instrumentation tool, and allocate and initialize mask and secrecy buffers.

\section{Evaluation}
\label{sec:evaluation}
We now evaluate the performance and security of the different \cc{} variants. We analyze whether there is remaining leakage with regard to 
collision attacks, and discuss trade-offs between security and performance.

\subsection{Experimental Setup}
We evaluate our proof-of-concept implementation of \cc{} against a number of typical algorithms which are used in widespread protocols like TLS or SSH. To observe variations caused by different implementations of the same primitive, we spread our analysis over several common libraries, that are \openssl{} 3.0.2, \wolfssl{} 5.3.0, \mbedtls{} 3.3.0 and \lsodium{} 1.0.18. 
As primitives which were shown to be vulnerable to ciphertext side-channel attacks~\cite{sytematicCipherleaks}, we picked EdDSA (Ed25519) and ECDSA (secp256r1), and verified that these are still vulnerable in the given implementations. \cite{DBLP:conf/uss/LiZWLC21} demonstrated an attack against the RSA signature scheme, which we included as well.
We also added ECDH (X25519) as a primitive that is widely used in cryptographic protocols and likely to be vulnerable as well. As additional benchmarks, we included the symmetric primitives AES-GCM and ChaCha20-Poly1305, the hash function SHA-512, and finally the Base64 decoding function as a non-cryptographic algorithm, that is nevertheless often present in cryptographic applications.

The analysis, instrumentation and all measurements were performed on an AMD EPYC 7763 CPU with Zen3 microarchitecture, which supports SEV-SNP. All libraries were compiled with GCC 9.4.0 on Ubuntu 20.04.4 LTS.
\texttt{MbedTLS} was linked statically, while the other libraries were linked as shared libraries.

\subsection{Performance}
\label{sec:eval:performance}
To get the information necessary for the mitigation, we ran the dynamic analysis as described in \Cref{sec:analysis}. We found that we achieve sufficient coverage by executing each target 10 times with random inputs in a loop, except for WolfSSL RSA, which required 20 due to high control flow variation introduced by blinding. In all cases, the time required for dynamic analysis was less than 5 minutes, with around 80\% of the time taken by the register tracking in the structure analysis, and most of the remaining time by the taint tracking. The most expensive target, \mbedtls{} ECDH, required tracking \num{170532009} executed instructions (\num{5167} unique). While the register tracking could be scrapped in favor of a faster but potentially less precise static liveness analysis (as done by several binary rewriting tools), note that these steps are executed offline and only need to be done once to protect a binary, so we deem an analysis time of a few minutes acceptable.

\subsubsection{Runtime overhead}
To measure the runtime overhead of the different \cc{} variants, we executed each target with \num{1000} random inputs, averaged the measured execution times, and computed the relative overhead compared to the original implementation. An overview of the resulting overall slowdowns of the different \cc{} variants is given in \Cref{table:runtime}.
As expected, \cffast{} has the lowest overhead, \cfenhanced{} has the highest, and \cfbase{} lies in between. The slowdown of \cfbase{} compared to \cffast{} is caused by the additional read for each protected memory access; in most cases, \cfenhanced{} performs quite similar to \cfbase{}, except for the symmetric primitives and utility functions which have a vastly higher number of 1-byte writes.

Moreover, generating masks with \rdr{} introduces a much higher overhead than with one of the other PRNGs.
This is caused by the continuous reseeding of the underlying shared hardware PRNG, in combination with \rdr{} not being designed for sampling random numbers at a high frequency.
The smallest overhead is achieved with the \aes{} PRNG, as it consists of a single \texttt{vaesenc} instruction and only needs two vector registers.
A detailed overview over all runtime overhead measurements is given in \Cref{tab:eval-overhead}.

%
\begingroup
\renewcommand{\arraystretch}{1.1}
\begin{table*}[h!t]
    \caption{Runtime overhead of instrumented binaries. For each \cc{} variant and PRNG, we measured the execution time in milliseconds (ms) of \num{1000} executions of each primitive and the corresponding overhead factor to the original implementation. The target \texttt{AES} refers to AES-GCM, the target \texttt{CC20} to ChaCha20-Poly1305. The last row shows the geometric mean of the respective overheads for each \cc{} variant.}
    \label{tab:eval-overhead}
    \centering
    \begin{tabular}{@{}b{2pt} p{22pt} l @{\hspace*{1em}}r @{\hspace*{2.5em}}r rr @{\hspace*{2.5em}}r rr @{\hspace*{2.5em}}r rr@{}}
        \toprule 
         & \multirow{2}[3]{*}{Target} & & \multirow{2}[3]{*}{\texttt{orig}} & \multicolumn{3}{c}{\hspace*{-2em}\textsc{CF-Fast}} & \multicolumn{3}{c}{\hspace*{-2em}\textsc{CF-Base}} & \multicolumn{3}{c}{\hspace*{0.5em}\textsc{CF-Enhanced}} \\
        \cmidrule(lr{2.5em}){5-7}
        \cmidrule(lr{2.5em}){8-10}
        \cmidrule(lr{-0.5pt}){11-13}
         & & & & \texttt{AES} & \texttt{XS\raisebox{.15ex}{\small +}} & \texttt{rdrand} & \texttt{AES} & \texttt{XS\raisebox{.15ex}{\small +}} & \texttt{rdrand} & \texttt{AES} & \texttt{XS\raisebox{.15ex}{\small +}} & \texttt{rdrand} \\
        \midrule
        
        \multirow{4}{*}{\rotatebox[origin=c]{90}{\textsf{libsodium}\hspace*{-1em}}} & \texttt{EdDSA} & time & \num{29} & \num{159} & \num{166} & \num{1159} & \num{189} & \num{248} & \num{1133} & \num{214} & \num{245} & \num{1134} \\
        & & factor & - & 5.5x & 5.7x & 40.0x & 6.5x & 8.6x & 39.1x & 7.4x & 8.4x & 39.1x \\[0.5em]

        & \texttt{SHA512} & time & \num{9} & \num{14} & \num{20} & \num{196} & \num{21} & \num{22} & \num{194} & \num{22} & \num{25} & \num{194} \\
        & & factor & - & 1.6x & 2.2x & 21.8x & 2.3x & 2.4x & 21.6x & 2.4x & 2.8x & 21.6x \\

        \midrule
        \multirow{3}{*}{\rotatebox[origin=c]{90}{\textsf{mbedTLS}}} & \texttt{AES} & time & \num{104} & \num{297} & \num{377} & \num{2849} & \num{364} & \num{371} & \num{2576} & \num{1204} & \num{1213} & \num{2683} \\
        & & factor & - & 2.9x & 3.6x & 27.4x & 3.5x & 3.6x & 24.8x & 11.6x & 11.7x & 25.8x \\[0.5em]

        & \texttt{Base64} & time & \num{10} & \num{12} & \num{13} & \num{58} & \num{16} & \num{16} & \num{45} & \num{28} & \num{30} & \num{46} \\
        & & factor & - & 1.2x & 1.3x & 5.8x & 1.6x & 1.6x & 4.5x & 2.8x & 3.0x & 4.6x \\[0.25em]

        & \texttt{CC20} & time & \num{144} & \num{324} & \num{332} & \num{2945} & \num{542} & \num{570} & \num{2952} & \num{1785} & \num{1721} & \num{3059} \\
        & & factor & - & 2.3x & 2.3x & 20.5x & 3.8x & 4.0x & 20.5x & 12.4x & 12.0x & 21.2x \\[0.5em]

        & \texttt{ECDH} & time & \num{1855} & \num{3674} & \num{3778} & \num{8559} & \num{9425} & \num{9440} & \num{14419} & \num{9926} & \num{10208} & \num{14827} \\
        & & factor & - & 2.0x & 2.0x & 4.6x & 5.1x & 5.1x & 7.8x & 5.4x & 5.5x & 8.0x \\[0.5em]

        & \texttt{ECDSA} & time & \num{472} & \num{3367} & \num{3558} & \num{8920} & \num{3912} & \num{3929} & \num{8297} & \num{4265} & \num{4301} & \num{8374} \\
        & & factor & - & 7.1x & 7.5x & 18.9x & 8.3x & 8.3x & 17.6x & 9.0x & 9.1x & 17.7x \\[0.5em]

        & \texttt{RSA} & time & \num{896} & \num{3276} & \num{3777} & \num{28886} & \num{5436} & \num{5339} & \num{27148} & \num{5527} & \num{5663} & \num{27208} \\
        & & factor & - & 3.7x & 4.2x & 32.2x & 6.1x & 6.0x & 30.3x & 6.2x & 6.3x & 30.4x \\

        \midrule
        \multirow{3}{*}{\rotatebox[origin=c]{90}{\textsf{OpenSSL}}} & \texttt{ECDH} & time & \num{172} & \num{541} & \num{550} & \num{2408} & \num{664} & \num{657} & \num{2323} & \num{708} & \num{807} & \num{2369} \\
        & & factor & - & 3.1x & 3.2x & 14.0x & 3.9x & 3.8x & 13.5x & 4.1x & 4.7x & 13.8x \\[0.5em]

        & \texttt{ECDSA} & time & \num{516} & \num{939} & \num{1121} & \num{7181} & \num{1795} & \num{1855} & \num{9980} & \num{2051} & \num{1973} & \num{10072} \\
        & & factor & - & 1.8x & 2.2x & 13.9x & 3.5x & 3.6x & 19.3x & 4.0x & 3.8x & 19.2x \\

        
        \midrule
        \multirow{3}{*}{\rotatebox[origin=c]{90}{\textsf{WolfSSL}}} & \texttt{AES} & time & \num{147} & \num{268} & \num{269} & \num{793} & \num{400} & \num{403} & \num{880} & \num{397} & \num{402} & \num{879} \\
        & & factor & - & 1.8x & 1.8x & 5.4x & 2.7x & 2.7x & 6.0x & 2.7x & 2.7x & 6.0x \\[0.5em]

        & \texttt{CC20} & time & \num{167} & \num{428} & \num{432} & \num{2787} & \num{596} & \num{630} & \num{2802} & \num{1157} & \num{1242} & \num{2874} \\
        & & factor & - & 2.6x & 2.6x & 16.7x & 3.6x & 3.8x & 16.8x & 6.9x & 7.4x & 17.2x \\[0.5em]

        & \texttt{ECDH} & time & \num{146} & \num{258} & \num{437} & \num{4217} & \num{544} & \num{565} & \num{4070} & \num{541} & \num{558} & \num{4070} \\
        & & factor & - & 1.8x & 3.0x & 28.9x & 3.7x & 3.9x & 27.9x & 3.7x & 3.8x & 27.9x \\[0.5em]

        & \texttt{ECDSA} & time & \num{1092} & \num{1704} & \num{1954} & \num{15765} & \num{3945} & \num{3834} & \num{18631} & \num{3883} & \num{3897} & \num{19654} \\
        & & factor & - & 1.6x & 1.8x & 14.4x & 3.6x & 3.5x & 17.1x & 3.6x & 3.6x & 17.1x \\[0.5em]

        & \texttt{EdDSA} & time & \num{60} & \num{124} & \num{156} & \num{1897} & \num{279} & \num{265} & \num{1759} & \num{280} & \num{290} & \num{1761} \\
        & & factor & - & 2.1x & 2.6x & 31.6x & 4.7x & 4.4x & 29.3x & 4.7x & 4.8x & 29.4x \\[0.5em]

        & \texttt{RSA} & time & \num{133} & \num{248} & \num{334} & \num{2901} & \num{588} & \num{605} & \num{2863} & \num{602} & \num{651} & \num{2870} \\
        & & factor & - & 1.9x & 2.5x & 21.8x & 4.4x & 4.5x & 21.5x & 4.5x & 4.9x & 21.6x \\

        \midrule
        & \multicolumn{2}{l}{average factor}  & - & \textbf{2.4x} & \textbf{2.7x} & \textbf{16.8x} & \textbf{3.9x} & \textbf{4.0x} & \textbf{17.3x} & \textbf{5.1x} & \textbf{5.3x} & \textbf{17.5x} \\
        
        \bottomrule
    \end{tabular}
\end{table*}
\endgroup

\subsubsection{Code properties contributing to overhead}
We identified several major factors that determine the overhead when hardening a particular implementation with \cc{}.
First of all, code that heavily relies on memory accesses for dealing with secret information is clearly more susceptible to overhead introduced by instrumentation than code that performs most computations in registers. This becomes apparent when comparing the RSA implementations of \mbedtls{} and \wolfssl{}: Though for \wolfssl{} a higher percentage of the memory accesses is instrumented (78\% writes vs. 65\%), \mbedtls{} has an order of magnitude more memory operations than \wolfssl{} %
 and thus gets a higher overhead.
Similarly, some instructions are more expensive than others in terms of ciphertext side-channel hardening: For example, arithmetic directly applied to memory operands requires a full decoding and re-encoding cycle (cf. \Cref{fig:instrumented-shr}), which is slow due to direct data dependencies between the steps. We observed this for \mbedtls{} ECDSA, which gets a much higher overhead (7.1x vs. 1.6x) than the comparable implementation in \wolfssl{}, mostly due to expensive \texttt{add}s in a hot code path.

Finally, the overhead is influenced by the general structure of the instrumented code, and the optimization capabilities of the binary rewriting framework. A framework operating at basic block level could perform better than our proof-of-concept implementation, which instruments each instruction in isolation to ease leakage analysis and debugging. For example, scratch registers may not need to be restored between usages, and instructions could be reordered to avoid saving status flags. This is particularly relevant as the compiler tends to interleave arithmetic instructions that have direct status flag dependencies with memory accesses (e.g., \texttt{add}-\texttt{mov}-\texttt{adc}). %

A detailed overview over the observed memory accesses is given in \Cref{tab:eval-writes} in \Cref{sec:appendix:eval}.

\begin{table}[t]
    \caption{Performance measurements for the different \cc{} variants and PRNGs.
    Each entry shows the geometric mean of the runtime overhead over all targets compared to the original, uninstrumented binary.}
    \centering
    \begin{tabular}{@{}l @{\hspace*{2em}}r @{\hspace*{2em}}r r@{}}
        \toprule
        & \aes{} & \texttt{XS\raisebox{.15ex}{\small +}} & \rdr{} \\
        \midrule
        \textsc{Fast} & 2.4x & 2.7x & 16.8x \\
        \textsc{Base} & 3.9x & 4.0x & 17.3x \\
        \textsc{Enhanced} & 5.1x & 5.3x & 17.5x \\
        \bottomrule
    \end{tabular}
    \label{table:runtime}
\end{table}

\subsection{Security}
\label{sec:eval:security}
In the following, we illustrate reasons for remaining collisions after applying the different variations of \cc{} and evaluate its practical security.

\subsubsection{Leakage sources}
\label{sec:eval:leakage-sources}
As we assume full path coverage of the implementation (see \Cref{sec:discussion:non-constant-time}) and our taint tracking does not undertaint, all vulnerable instructions are identified and protected. Thus, the only remaining source of leakage are \textbf{collisions of the masks or the masked plaintexts}: With \cfbase{}, the secrecy information is stored in a separate buffer. If the mask $M$ is fully random and independent from the plaintext $P$, the masked plaintext $\hat{P}$ becomes independent from $P$ as well.
However, the attacker can access both ciphertexts $C_{\hat{P}}=\encPt(\hat{P})$ and $C_M=\encMask(M)$, so they are able to detect whether $\hat{P}$ or $M$ appear repeatedly.
If the data memory block is rarely changed and the number of protected bits is sufficiently low, a mask collision is possible and may leak information about the plaintext. %
A similar issue can occur with \cffast{}, which stores the secrecy information directly in the mask buffer by setting the mask to zero for public values. We can assume that the attacker knows the ciphertext $C_0=\encPt(0)$ of an unmasked zeroed data block, as memory usually is zero initialized. If they observe $C_0$ again after a write of $P$ with mask $M\neq 0$, they can use that $C_0=\encPt(P\oplus M)$ and thus $P=M$ to infer that $P\neq 0$.
These leakages through masks or masked plaintexts are mostly relevant for 1-byte writes to variables in memory blocks with little other activity. With \cfenhanced{}, we enforce a minimum width of masked data, which further reduces the probability of mask collisions and other non-unique writes at the cost of a slightly higher overhead.

Another factor is the \textbf{quality of the PRNG used for mask generation}, for which we identified two primary criteria. First, the pseudorandomness should not correlate with the plaintexts: For example, simply incrementing the masks may lead to many collisions of the masked plaintexts in algorithms that use linear arithmetic. Second, deterministic PRNGs should have a sufficient cycle length, to keep an attacker from reliably triggering the same mask at the same address during the application's runtime. \texttt{Rdrand} offers the fastest available solution for cryptographically secure pseudorandomness. However, given the subsequent memory encryption, the used PRNG does not necessarily need to be cryptographic, as long as it satisfies the above criteria and thus does not tend to generate repeating masks or masked plaintexts. \xsp{} has a cycle length of $2^{128}-1$ and passes all \emph{BigCrush} tests of the \emph{TestU01} suite~\cite{DBLP:journals/toms/LEcuyerS07}, though it has some weaknesses~\cite{DBLP:journals/jcam/HaramotoMS22}.
Our custom one-round \aes{} PRNG passes all \emph{BigCrush} tests and seems to perform well in practice, but does not have a guaranteed cycle length. We leave this analysis
to future work.

\subsubsection{Observed collisions}
\label{sec:eval:leakage-analysis}

To analyze potentially remaining ciphertext collisions, we extended the taint tracking to export a full trace of all memory writes alongside corresponding secrecy information. We then created a \pintool{} that generates a trace of all memory writes for an instrumented binary.
As each original memory access may be replaced by multiple memory accesses during instrumentation, we inserted special marker instructions that denote the beginning and end of a particular instrumented memory access sequence.
With this information, we align the traces using a custom evaluation tool, and proceed with checking whether there are repeated writes of the same secret value to the same address.
As the dictionary attack builds upon the collision attack, finding no collisions implies security against all known ciphertext side-channel attack primitives.
We found that using the same amount of test cases for our evaluation as for the initial taint tracking was sufficient, as due to the size and complexity of the evaluated targets systematic issues already appear during the first few executions.

We were able to confirm the suspected remaining leakages with our evaluation. For example, there are several thousand collisions for \cfbase{} and \cffast{} with the \mbedtls{} AES-GCM target, which encrypts 16 KiB of plaintext using AES-NI and has \num{812120} 1-byte writes, which is 66\% of its total writes. The observed collisions both included repeating masks and cases where applying a new mask to a new plaintext led to the same result. All colliding 1-byte writes were related to sequential writing into an array, e.g., when data is copied or buffers are cleared between different processing steps. The corresponding collisions had high temporal locality and the respective 16-byte blocks only appeared exactly two times, so while there is some leakage, its exploitability is limited. With \cfenhanced{}, all collisions disappeared. All observed collisions in the analyzed targets were for 1-byte writes, which suggests that restricting \cfenhanced{} to 1-byte writes (and possibly 2-byte writes) is sufficient. We encountered almost no 2-byte writes in our experiments. We further discuss the security impact of the collisions in \cffast{} in \Cref{sec:eval:tradeoff}.

We did not see any relevant difference between the particular PRNGs: The number of collisions is roughly equal, and there was no 32-bit mask collision even for the targets with the highest number of instrumented writes. This suggests that they are all generally suited for generating masks for the evaluated primitives within the given constraints. Nevertheless, the decision for a particular PRNG should not be made easily, as is discussed in the next section. %

\subsection{Balancing Security and Performance}
\label{sec:eval:tradeoff}
Each variant and PRNG comes with its own advantages and drawbacks. We point out some guidelines for choosing the best composition for a given use case.

\subsubsection{Properties of the implementation}
To determine the most suitable variant of \cc{}, one should look at the properties of the given implementation. 
For example, symmetric primitives, which showed a huge amount of 1-byte writes in our evaluation, do not necessarily need to be hardened against ciphertext side-channels.
With hardware extensions like AES-NI and CLMUL, we found that leakage is mostly restricted to copying of inputs and outputs between encryption rounds. Thus, if the same buffer is reused for multiple blocks, the attacker may occasionally learn that a particular plaintext block or parts of it repeat.
Whether this is tolerable depends on the specific use case.

\subsubsection{Choosing a \cc{} variant}
While \cffast{} has the least performance overhead, it has the additional risk of leaking whether the mask and the plaintext are equal, as described in \Cref{sec:eval:leakage-sources}. While we did not observe that particular scenario, we saw several 1-byte collisions in \wolfssl{}'s X25519 \texttt{cswap} implementation. This suggests that \cffast{} and \cfbase{} are dangerous even for algorithms with a very small number of 1-byte writes.
Future work may develop a further variant that uses a merged mask/secrecy buffer but widens small writes to 4 bytes, to get both the performance benefit of \cffast{} and the protection of \cfenhanced{}.
For deciding between \cfbase{} and \cfenhanced{}, we generally recommend choosing the latter due to the better protection of 1-byte writes. While we observed a higher performance overhead, the difference was almost exclusively caused by the symmetric primitives which do a lot of 1-byte operations. Excluding the symmetric implementations from the geometric mean yields an overhead of 5.2x for \cfenhanced{} versus 4.9x for \cfbase{}
and the \xsp{} PRNG.

\subsubsection{Choosing a PRNG}
Despite the high security guarantees, the considerable performance overhead of \cc{} with \rdr{} suggests that this PRNG is not suitable for use with primitives that have a lot of vulnerable memory accesses. On the other hand, our custom \aes{} PRNG is very fast and did not exhibit more collisions than the other PRNGs in our experiments, but is not well examined in terms of statistical properties and cycle length. Thus, as a compromise, we suggest using a fast PRNG that is well-analyzed and meets the criteria outlined in \Cref{sec:eval:leakage-sources}, such as \xsp{}, which only introduced a slightly higher overhead than \aes{}.
As a workaround for an insufficient cycle length or concerns that a high number of samples may expose weaknesses, the PRNG may be periodically reseeded with fresh entropy via instructions like \texttt{rdseed}, e.g., each time before the hardened primitive is executed.
Finally, a production-level implementation of \cc{} may combine different PRNGs, like a fast one for hot code paths and \rdr{} elsewhere.

\subsubsection{Practical impact of overhead}
Note that we focused our performance analysis on isolated cryptographic primitives, which does not reflect their typical use case. Instead, they are usually embedded into a higher-level application like a network protocol, which limits the practical influence of a moderate overhead in a specific component. For example, in TLS, only the handshake is subject to asymmetric cryptography that needs to be hardened against ciphertext side-channel attacks. The predominant part of the protocol's runtime, the symmetric encryption and transmission of the payload, may not need as much costly protection.%

\section{Discussion}
\label{sec:discussion}
We conclude our study with a discussion of some design decisions of \cc{}, and point out possible angles for future work which may improve accuracy and performance.

\subsection{Source Code vs. Binary Instrumentation}
\label{sec:discussion:code-based-instrumentation}

Instead of instrumenting binaries, the implementations could be hardened during compilation:
As the compiler can freely adapt the code layout and is not restricted during register allocation, it can generate more efficient binaries.
However, this comes with some obstacles. First, a source-based approach would need to be able to deal with handwritten assembly code, which is abundant in highly-optimized libraries like OpenSSL or libc. This assembly code is opaque to the compiler, but can be handled transparently by binary instrumentation.

A second obstacle is a leakage analysis that spans multiple libraries. At the beginning, the application developer would need to checkout the source code of all relevant dependencies, such that they can be recompiled with the appropriate protection. The compiler can then conduct a static data flow analysis that identifies all program points that may come in contact with secret data~\cite{DBLP:conf/ccs/BorrelloDQG21}. As we found during our experiments, a particular library may call a function in another library with secret parameters, so conducting a leakage analysis on a library in isolation is insufficient. This leaves two options: First, the leakage analysis can choose to protect the parameters of the entire outward facing API of a given library, such that all incoming function calls are assumed as passing secret data. However, this significant overapproximation is likely to neutralize the performance benefit of a compiler-based solution. Thus, as a second option, we may try to conduct the leakage analysis over all code bases at once.
This is hindered by the fact that static analysis of a large code base like OpenSSL or libc is already difficult, and even more so when looking at several such code bases with different build systems and structure. At the very least, it would require lots of manual tuning by the application developer.

An alternative to binary rewriting that is worth exploring for a production-level implementation of \cc{} is a hybrid approach combining dynamic analysis and compiler-based instrumentation: First, a dynamic analysis is conducted over all libraries as described in \Cref{sec:analysis}. However, the results are then not used to instrument the binaries using SBI, but are sent back to the compiler. A suitable level for this is the intermediate representation (IR) of LLVM: The IR can be executed through a VM, enabling dynamic analysis. At the same time, it is abstract enough to still allow compiler optimizations between inserting the masking code and generating ELF binaries. Applying the analysis and instrumentation to IR also avoids the practical problems of dealing with large code bases, as those can be normally translated and linked into IR files. However, contrary to binary rewriting, this method still requires some effort from the library developer, and cannot straightforwardly deal with handwritten assembly code, that would need to be lifted to an equivalent IR representation first. Finally, advanced binary rewriting engines that generate symbolized reassemblable disassembly already offer performance similar to the compiler.

\subsection{Analysis Coverage}
\label{sec:discussion:non-constant-time}

Independent of the approach on instrumentation, we need to find all loads and stores that ever deal with protected data.
Missing instructions during the secrecy analysis may lead to loading or storing invalid data, which can in turn cause functional incorrectness or crashes of the hardened binary.
In constant-time implementations, there are no secret-dependent branches and memory accesses.
However, it is useful to support some secret-independent control flow variation, e.g., for error handling or processing messages of varying length.
As our analysis is dynamic, we have to rely on our inputs generating sufficient coverage, that is covering every possible execution path between classification and declassification of secrets. The secret tracking must not underapproximate (undertaint), as this may lead to missing leakages or instability due to instructions that cannot handle masked data. Overapproximation (overtainting) is acceptable to speed up leakage analysis, but may lead to unnecessary instrumentation and thus a higher runtime overhead.
We found that few random inputs were sufficient to get the coverage needed for our analysis; however, one could also employ techniques like fuzzing to maximize the chances of finding all relevant code paths, especially when applying \cc{} to non constant-time code.
Fuzzing and a larger test case body would only impact the overhead of the offline analysis step.
Another approach for achieving full coverage is using a purely static analysis, which may be conducted either on binaries or as part of a pure compiler-based solution.
However, even for the smaller exploitable primitives, we measured several tens of millions of executed instructions for a single dynamic analysis iteration, which poses a huge amount of instructions to analyze for a static analysis. To make this feasible, the static analysis would need to make some approximations, which would in turn increase the runtime overhead of the mitigation.%

\subsection{Alternatives to Masking}
\label{sec:discussion:masking-alternatives}
Our masking approach ensures that the written values are independent from the actual plaintexts.
However, as mentioned in~\cite{sytematicCipherleaks}, instead of randomizing the values written to the same address, it is also possible to randomize the address itself.
This approach would need a separate memory area for secret data. The area can, for example, be implemented as a queue with used and free space that is updated with each write. The original memory locations then point to the corresponding block in the secure memory area. The resulting memory overhead becomes a security parameter: The bigger the secure memory area, the lower the risk of collisions. In early experiments, we found that the instrumentation for this approach would have significantly higher overhead due to the management of the queue. It is better suited for narrow cases where code that deals with a well-defined data structure is hardened manually, e.g., the register save/restore during a kernel context switch. In our setting, we do not see an advantage of using randomized addresses instead of masking.

In a compiler-based setting, it is also possible to securely store data by interleaving it with random nonces. For example, each 16-byte block in AMD SEV can be split into two 8-byte halves, where the first half receives the payload, while the second half is treated as a nonce that is incremented on each write. Note that this has to be done in a single step, so the entire block may need to be buffered in a vector register, that is then written at once. This method guarantees that there are no collisions for $2^{64}$ writes to a given block, and has a higher locality of memory accesses, as no mask buffer is necessary. In addition, reads are almost as fast as for unprotected data, as no decoding is necessary. However, it has a high implementation complexity, as the compiler has to detect code that uses pointers to iterate over arrays and adjust such loops accordingly. Finally, the compiler needs to install logic for detecting unaligned accesses that may span multiple payload blocks, introducing a different kind of overhead. Nevertheless, interleaving may be worth exploring for programming languages that abstract away the memory layout of data structures and do not allow raw pointers.

\subsection{Compatibility to CFI}
\label{sec:discussion:compatibility-to-cfi}
Along with constant-time code and ciphertext side-channel mitigations, there are further mechanisms for ensuring secure code execution, an important one being control flow integrity (CFI) protection. For example, Intel and AMD provide the so-called control flow enforcement technology (CET), that detects unwanted control flow modifications through a shadow stack and by enforcing that indirect jumps and calls point to special \texttt{endbr64} instructions.
Besides inserting direct jumps to the instrumentation section, which may be avoided by using a more sophisticated binary rewriting framework, our ciphertext side-channel mitigation does not modify the control flow. Indirect branches still point to \texttt{endbr64} instructions, and the call stack is left untouched. Thus, \cc{} is compatible with CFI mechanisms like CET.

\section{Related Work}
\label{sec:relaated-work}

\bheading{Dynamic taint analysis} is a software analysis technique that is implemented in a variety of tools~\cite{DBLP:conf/vee/KemerlisPJK12,DBLP:conf/ndss/ChuaWBSLS19,DBLP:conf/raid/DavanianQQY19,DBLP:conf/issta/ClauseLO07,DBLP:conf/micro/QinWLKZW06,DBLP:conf/ndss/KangMPS11}.
Data flow based information tracking can support finding vulnerabilities in source or binary code.
On the one hand, it can be used to increase the branch coverage of fuzzers like the AngoraFuzzer~\cite{DBLP:conf/sp/ChenC18} or VUzzer~\cite{DBLP:conf/ndss/0001JKCGB17} by checking on which bytes of secret inputs branching decisions are based. %
On the other hand, taint analysis can help to keep sensitive data always encrypted in memory through data protection tools like DynPTA~\cite{DBLP:conf/sp/PalitMMP21} which is a compiler-based approach.%

\bheading{Automated analysis of side-channels} in binaries focuses on finding non-constant-time behavior by analyzing leakages that can be modeled in different ways.
There is a number of tools, which use DBI to observe leakages at runtime~\cite{DBLP:conf/uss/WeiserZSMMS18,DBLP:conf/ccs/WichelmannSP022} or detect secret-dependent accesses through symbolic execution~\cite{DBLP:conf/uss/WangWLZW17,DBLP:conf/uss/0011BL0ZW19,DBLP:journals/acm/DanielBR22}.
Those existing tools for finding side-channel leakages do not cover the ciphertext side-channel attack vector, as it is not originated from a deviation in the behavior of memory accesses, but rather from the content of write accesses which affects the ciphertexts.
However, they can be used to initially verify whether the code is constant-time, as non-constant-time code is even easier to attack than through the ciphertext side-channel.

\bheading{Memory protection mechanisms} implement the protection of sensitive data in memory.
\emph{Data space randomization} (DSR)~\cite{DBLP:conf/dimva/BhatkarS08} randomizes the representation of data that is stored in memory, with the aim of thwarting control flow hijacking attacks.
This is done by instrumenting the code so that masks are added to or removed from variables before or after memory load and store operations.
\emph{CoDaRR}~\cite{DBLP:conf/ccs/RajasekaranCGNV20} extends DSR with a protection against leaking the masks that are used for DSR so that rerandomization prevents from recovering the secrets through attacking masks.
These solutions are source code-based and thus not applicable for our tool.

\bheading{Static binary instrumentation} builds the basis for binary-level analysis and protection tools with different ways to insert additional code.
The trampoline SBI approach is used by tools like \emph{Detours}~\cite{hunt1999detours} and \emph{PEBIL}~\cite{DBLP:conf/ispass/LaurenzanoTCS10} which relocate functions to newly-added \texttt{.text} and \texttt{.data} sections together with a redirection to these sections through 5-byte jumps.
The technique is extended with inserting \texttt{int3} when a jump instruction does not fit %
in \emph{BIRD}~\cite{DBLP:conf/cgo/NandaLLC06} and short 2-byte intermediate jumps in \emph{DynInst}~\cite{DBLP:journals/ijhpca/BuckH00,DBLP:conf/IEEEpact/HollingsworthMGNXZ97}.
In our work, we implemented an optimized combination of different jumps and \texttt{int3} to build a lightweight static instrumentation.
For a production-level implementation of \cc{}, a sophisticated instrumentation framework should be used, but for our study, a custom tool
tailored to the interaction with the dynamic analyses
was easier integrated.
Another way of coping with 5-byte jumps is \emph{instruction punning}, as implemented in \emph{LiteInst}~\cite{DBLP:conf/pldi/ChamithSDN17} and \emph{E9PATCH}~\cite{DBLP:conf/pldi/DuckGR20}.
This technique uses address offset bytes in a jump instruction to also encode instructions, so fewer bytes need to be overwritten.
For our mitigation implementation, we did not employ instruction punning, as it introduces additional complexity and memory overhead due to the jump targets being scattered over a large memory area.
\emph{RetroWrite}~\cite{DBLP:conf/sp/DineshBXP20} uses symbolization to generate reassemblable assembly that can be equipped with instrumentation passes and yields an optimized instrumented binary.
Layout-agnostic binary rewriting can be performed with \emph{Egalito}~\cite{DBLP:conf/asplos/Williams-KingKW20} that uses metadata to lift the program into a specialized intermediate representation.
These approaches yield more efficient binaries, but need additional support for stripped binaries and some forms of inline assembly as used by libraries like OpenSSL, respectively.

\section{Conclusion} \label{sec:conclusion}
In this work, we have presented a drop-in technique for automatically protecting binaries from leaking processed secrets through a ciphertext side-channel.
Our approach comprises finding vulnerable code parts and then protecting them
by preventing observable ciphertext changes based on secret data.
The leakage localization technique combines dynamic binary instrumentation and dynamic taint analysis to protect only those memory accesses that deal with secrets or secret-derived data.
The mitigation introduces randomness such that the plaintexts written to memory change for each write, leading to corresponding unique ciphertexts.
We have shown that the highest security level of our proof-of-concept implementation can detect and mitigate all leaking memory accesses, with a very small probability of remaining leakage.
Since there is no indication of fixes for existing or upcoming hardware, \cc{} is a suitable approach for protecting software against the ciphertext side-channel.

\section*{Acknowledgements}
We would like to thank Gregor Leander for a helpful discussion on the security of fast PRNGs, and the anonymous reviewers and our shepherd for their detailed comments and suggestions for improvement.
This work has been supported by Deutsche Forschungsgemeinschaft (DFG) under grants 427774779 and 439797619, and by Bundesministerium für Bildung und Forschung (BMBF) through the ENCOPIA and SASVI projects.

\ifUsenix
  \bibliographystyle{plain}
  {\footnotesize
  \bibliography{references-short}
  }
\else
  \bibliographystyle{IEEEtranS}
  {\footnotesize
  \bibliography{IEEEabrv,references}
  }
\fi

\onecolumn
\begin{appendix}

\section{Static Instrumentation Example}
%
%
%
\begin{figure*}[h]
    \centering
    \includegraphics[height=12.8em]{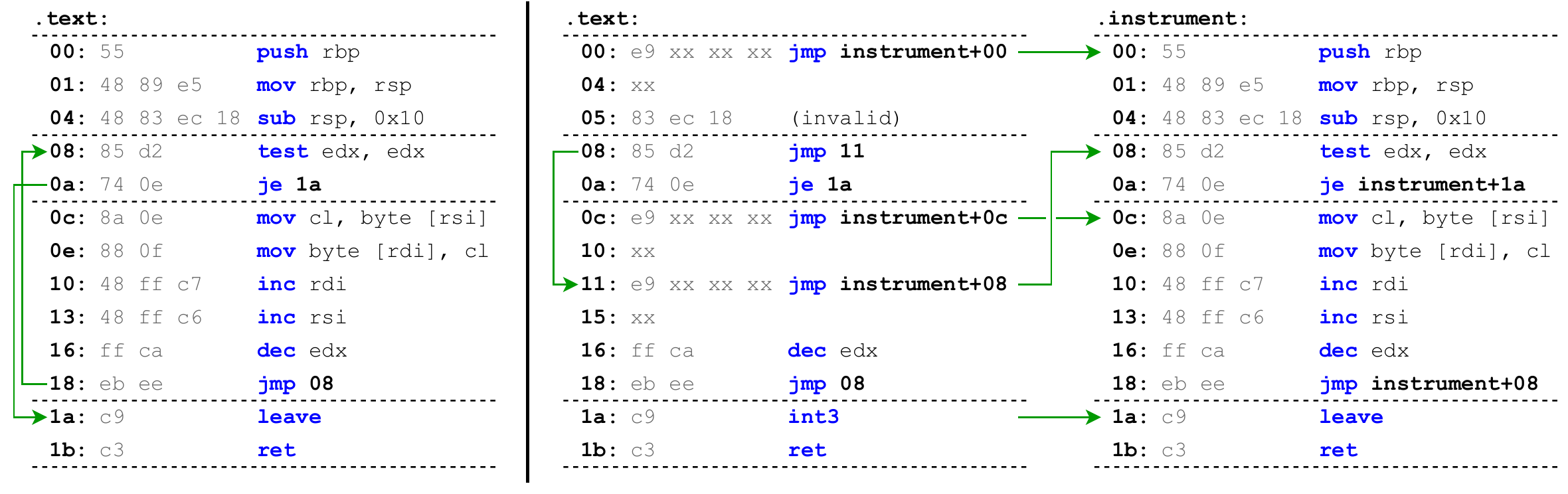}
    \caption{A simple \texttt{memcpy} implementation (left), and the resulting static instrumentation (right). The basic blocks of the original code are separated by dashed lines, control flow edges are marked with arrows. The first basic block has sufficient space for a direct 5-byte jump to the instrumentation code. The second basic block only has 4 bytes, but the third basic block offers space for two 5-byte jumps, so the second basic block gets a 2-byte jump to the third basic block (offset \texttt{11}) and from there a 5-byte jump to the instrumentation code. For the fourth basic block, all remaining space in the other basic blocks is already consumed, so it has to use an \texttt{int3} instruction. Execution that ends up at the beginning of any of the original basic blocks is always redirected to their counterparts in the \texttt{.instrument} section.}
    \label{fig:static-instrumentation}
\end{figure*}
\section{Evaluation Results}
\label{sec:appendix:eval}

\begingroup
\begin{table*}[h!]
    \caption{Memory accesses that have to be instrumented. Writes are split by their size, whereby \#$n$ denotes the number of $n$-byte writes. \% instr. reads/writes shows the respective total percentage of instrumented accesses. Each target was iterated 10 times.}
    \label{tab:eval-writes}
    \centering
    \begin{tabular}{@{\hspace*{0.2em}}l @{\hspace*{0.5em}}r r >{\bfseries}r *{7}{r} >{\bfseries}r @{}}
        \toprule
        \multirow{2}[3]{*}{\hspace*{-0.2em}Target} & \multicolumn{1}{r}{\multirow{2}[3]{*}{\# reads}} & \multicolumn{2}{c}{instr. reads} & \multicolumn{1}{r}{\multirow{2}[3]{*}{\# writes}} & \multicolumn{7}{c}{instrumented writes}  \\ 
        \cmidrule(lr){3-4}
        \cmidrule(lr{-0.5pt}){6-12}
         & & \# & \normalfont{\%} & & \#1 & \#2 & \#4 & \#8 & \#16 & \#32 & \normalfont{\%} \\
        \midrule
        \multicolumn{12}{l}{\hspace*{-0.5em}\underline{\lsodium{}}}\\
         \texttt{EdDSA} & \num{648453} & \num{448415} & 69 & \num{441736} & \num{4681} & \num{0} & \num{0} & \num{372600} & \num{6180} & \num{1160} & 87 \\
         \texttt{SHA512} & \num{200328} & \num{82722} & 41 & \num{104000} & \num{810} & \num{0} & \num{0} & \num{58718} & \num{4800} & \num{784} & 62 \\
        \midrule

        \multicolumn{12}{l}{\hspace*{-0.5em}\underline{\mbedtls{}}}\\
         \texttt{AES} & \num{1887551} & \num{1403255} & 74 & \num{1237457} & \num{812120} & \num{0} & \num{42} & \num{20715} & \num{30256} & \num{304} & 70 \\
         \texttt{Base64} & \num{195458} & \num{16020} & 8 & \num{128552} & \num{23599} & \num{0} & \num{0} & \num{5130} & \num{0} & \num{0} & 22 \\
         \texttt{CC20} & \num{1737111} & \num{1487956} & 86 & \num{1105221} & \num{641280} & \num{0} & \num{250910} & \num{217} & \num{60140} & \num{10068} & 87 \\
         \texttt{ECDH} & \num{37328410} & \num{3454726} & 9 & \num{18773246} & \num{0} & \num{0} & \num{881397} & \num{1566188} & \num{0} & \num{1172058} & 19 \\
         \texttt{ECDSA} & \num{7120602} & \num{3301437} & 46 & \num{3748086} & \num{14240} & \num{10} & \num{260673} & \num{1447753} & \num{7674} & \num{123806} & 49 \\
         \texttt{RSA} & \num{21203381} & \num{12012577} & 57 & \num{12068011} & \num{1950} & \num{10} & \num{360804} & \num{7303398} & \num{1243} & \num{122320} & 65 \\
        \midrule

        \multicolumn{12}{l}{\hspace*{-0.5em}\underline{\openssl{}}}\\
        \texttt{ECDH} & \num{4799917} & \num{390344} & 8 & \num{2532111} & \num{2750} & \num{0} & \num{2550} & \num{248691} & \num{62} & \num{470} & 10 \\
        \texttt{ECDSA} & \num{12041083} & \num{5463996} & 45 & \num{6950318} & \num{2329} & \num{0} & \num{524025} & \num{2671708} & \num{1492} & \num{3762} & 46 \\
        \midrule


        \multicolumn{12}{l}{\hspace*{-0.5em}\underline{\wolfssl{}}}\\
        \texttt{AES} & \num{3661782} & \num{1550484} & 42 & \num{288454} & \num{13150} & \num{0} & \num{90630} & \num{60427} & \num{10234} & \num{0} & 60 \\
        \texttt{CC20} & \num{2603432} & \num{1547406} & 59 & \num{994267} & \num{320320} & \num{0} & \num{476140} & \num{25893} & \num{20020} & \num{0} & 85 \\
        \texttt{ECDH} & \num{2317955} & \num{1753953} & 76 & \num{1916549} & \num{1248} & \num{0} & \num{10752} & \num{1409475} & \num{20} & \num{0} & 74 \\
        \texttt{ECDSA} & \num{19969154} & \num{11606148} & 58 & \num{9519250} & \num{721} & \num{0} & \num{543354} & \num{5292431} & \num{258140} & \num{1584} & 64 \\
        \texttt{EdDSA} & \num{1213466} & \num{694483} & 57 & \num{884122} & \num{4711} & \num{0} & \num{11560} & \num{568368} & \num{40} & \num{82} & 66 \\
        \texttt{RSA} & \num{2350077} & \num{1886260} & 80 & \num{1193204} & \num{1351} & \num{0} & \num{106801} & \num{753096} & \num{20580} & \num{46176} & 78 \\
        \bottomrule
    \end{tabular}
\end{table*}
\endgroup

\end{appendix}

\end{document}